\documentclass[twocolumn]{webofc}

\usepackage[varg]{txfonts}   
\begin{document}
\title{Search for the $\boldsymbol{N}$(1685) in $\boldsymbol{\eta\pi}$-Photoproduction}
%
%

\author{\firstname{Dominik} \lastname{Werthm\"uller}\inst{1}\fnsep\thanks{\email{dominik.werthmueller@york.ac.uk}}
        on behalf of the A2 Collaboration at MAMI
}

\institute{Department of Physics, University of York, Heslington, York, Y010 5DD, UK
          }

\abstract{%
The nucleon-like member $N(1685)$ of the speculative baryon
antidecuplet denotes one possible explanation for the narrow 
peak-structure around $W = 1.68$ GeV observed in the total cross
section of $\eta$-photoproduction off the neutron. If this baryon
existed, it would likely to be seen in other reactions as well.
While the aforementioned peak, whatever its nature is, was confirmed
by several experiments, claims for signatures of the $N(1685)$ in
other reactions and observables are mainly made by V. Kuznetsov
{\it et al.} using GRAAL data. Their latest work suggests signals of both
$N(1685)$ charge states in all isospin channels of $\eta\pi$-
photoproduction off the proton and neutron. This contribution
reports on challenging these claims with data from the A2 at MAMI
experiment employing photon beam energies from $E_{\gamma}$=1.43--1.58
GeV. The $\eta\pi^0p$ and $\eta\pi^+n$ final states produced from
a hydrogen target were studied and new analysis cuts were tested in
order to enhance a possible signal.
}
\maketitle
\section{Introduction}
\label{sec:intro}
The search for exotic states denotes a major focus
of contemporary hadron physics. Such bound states of
the strong interaction are not built from ordinary
$q\bar{q}$ (for mesons) and $qqq$ (for baryons)
configurations in terms of the quark model but
involve $q\bar{q}q\bar{q}$ (tetraquark),
$qqqq\bar{q}$ (pentaquark), etc.\ as well as gluonic degrees of freedom
to constitute a color singlet.

\subsection{Exotic Pentaquarks}

The most recent experimental evidence for pentaquark states
was claimed in 2015 by the LHCb collaboration
in the $J/\psi p$ final state of $\Lambda_b^0$ decays
hinting at an $uudc\bar{c}$
quark content \cite{Aaij:2015fs}.
In the meantime, a total of three states are thought to be
identified \citep{Aaij:2019}. GlueX, a first independent experiment
using $J/\psi$-photoproduction as an alternative
production mechanism could
not see any evidence of these pentaquark states \cite{GlueX_Jpsi}.
More data from GlueX and other experiments
are being taken and analyzed, and theoretical
efforts to understand the production mechanisms
are ongoing \cite{Winney_19}.

Before the evidence for pentaquarks containing heavy
charm quarks emerged, in 2003 the LEPS collaboration
claimed evidence for a $uudd\bar{s}$
pentaquark state named $\Theta^+$ \cite{LEPS_Pentaquark}.
In the peculiar episode that followed the announcement, several
experiments initially confirmed the existence of a structure
around ${W=1540}$ MeV in the $KN$ system. In the end,
however, large
statistics measurements, mainly from the CLAS experiment,
could not confirm the initial findings (see \cite{Hicks_Review} for
an overview). Nevertheless, as the detector acceptances of CLAS and 
LEPS are quite different, new LEPS results using an
improved setup and analysis 
are still awaited eagerly \cite{LEPS_Proceeding}.
In addition, there are newer positive claims from
the DIANA experiment \cite{DIANA_Pentaquark} and a sub-group
of the CLAS collaboration \cite{Amaryan:2012gf}.

\begin{figure}[h]
\centering
\includegraphics[width=0.8\columnwidth]{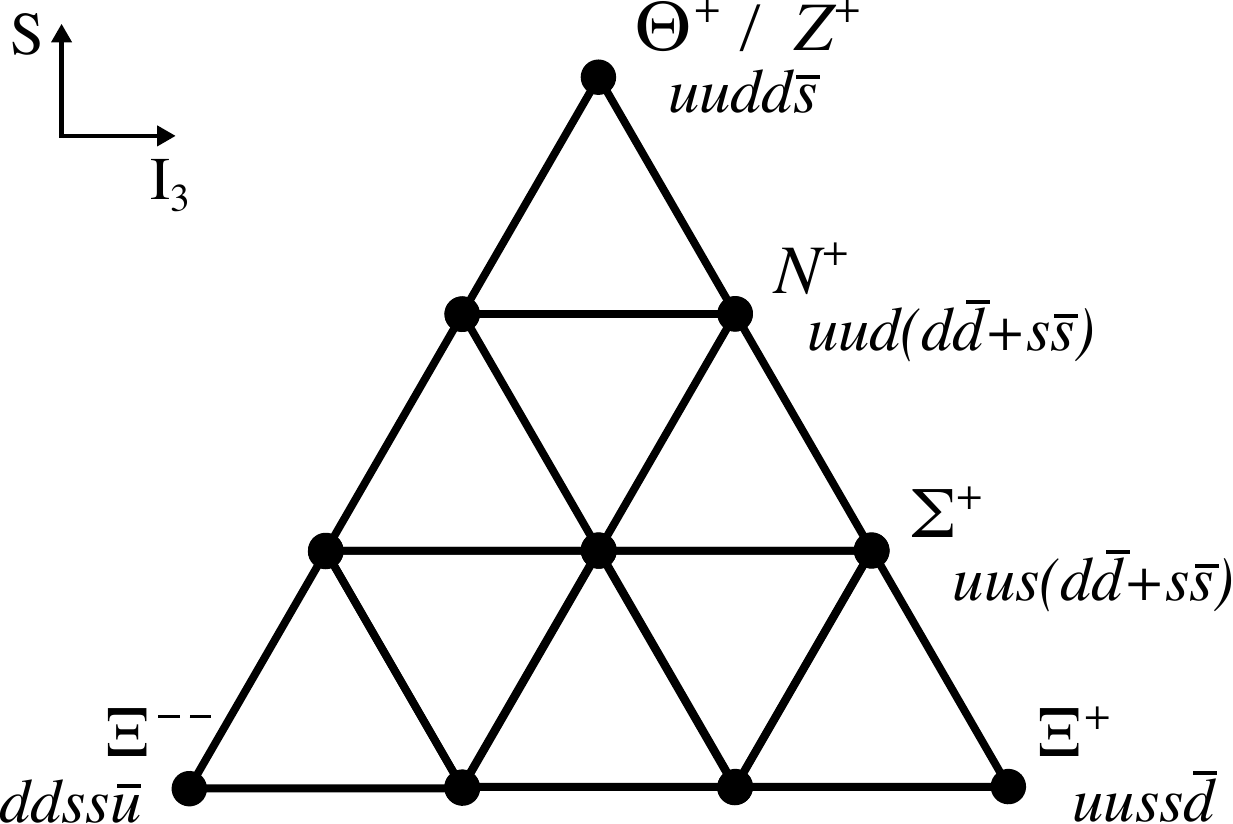}
\caption{Antidecuplet $\overline{\boldsymbol{10}}$ predicted by the Chiral Quark-Soliton Model
\cite{CQSM_Pred}.}
\label{fig:antidecuplet}
\end{figure}

Pentaquark states were studied using the quark model long before
the $\Theta^+$ claim \cite{Jaffe_Exotics, HOGAASEN1978119}.
The search for such a state at LEPS, although, 
was motivated by the striking
prediction of the Chiral Quark-Soliton Model
\cite{CQSM_Pred}, a Skyrme model in which baryons are described
as solitons of the chiral field. This model not only describes
the known octet and decuplet but also predicts the antidecuplet
depicted in figure~\ref{fig:antidecuplet}. The exotic $Z^+$
(later renamed to $\Theta^+$) with 
isospin $I = 0$ and strangeness $S = 1$ was predicted to be as
narrow as 1 MeV and to decay into $K^+n$ or $K^0p$, which matched
the LEPS results.

\subsection{The Cryptoexotic $\boldsymbol{N(1685)}$}

The nucleon-like antidecuplet state
$N_{\overline{10}}$ was first identified with the
nucleon resonance ${N(1710)1/2^+}$ (although
the width of the $N_{\overline{10}}$ was predicted
to be much more narrow) in order to calculate the masses
of the other members of the multiplet \cite{CQSM_Pred}.
The $N_{\overline{10}}$ (as well as the $\Sigma_{\overline{10}}$)
is cryptoexotic, i.e., its quantum numbers can also be realized
in an ordinary $qqq$ configuration.
It was quite a surprise when in 2007, Kuznetsov {\it et al.}
(members of the GRAAL collaboration)
presented evidence for a $N_{\overline{10}}$ candidate with
matching properties \cite{Kuznetsov_07}:
The bump at ${W=1685}$ MeV was observed in the
cross section of $\eta$-photoproduction on the neutron
 in agreement with
the leading $N_{\overline{10}}\to\eta N$ decay
\cite{CQSM_Pred}. Secondly, the
width of the structure was very narrow (around 30 MeV)
\cite{Arndt_04} compared to usual nucleon resonances.
Thirdly, the absence of any bump in the corresponding cross section
on the proton matched the predicted small photocoupling
compared to the neutron (`neutron anomaly') \cite{Polyakov_Rathke}.

While first there were doubts about the presence of the structure
in the $\eta$-photoproduction cross section, the bump was
observed by several independent experiments, such as
LNS \cite{Miyahara_07},
CBELSA/TAPS \cite{Jaegle_08,Jaegle_11} and
A2 at MAMI \cite{Werthmueller_13,Werthmueller_14,Witthauer_13}.
Similar structures
are also claimed to be present in different reaction channels
and observables, namely the beam asymmetry $\Sigma$ in
$\gamma p\to \eta p$ \cite{Kuznetsov2008}, and in 
Compton scattering off the neutron (cross section 
\cite{Kuznetsov_Compton_CS}) and off the proton
(beam asymmetry $\Sigma$ \cite{Kuznetsov_Compton_Sigma}).
These results all originate from the GRAAL experiment
and remain unconfirmed by any other experiment so far.

In view of some early analyses of the available data
supporting the scenario of a new narrow 
resonance \cite{Arndt_04,Choi_06,Fix_07,Anisovich_09,Shrestha_12},
a $N(1685)$ with unknown quantum numbers
was added as a one-star resonance to the 2012 Review
of Particle Physics \cite{PDG_2012}
by the Particle Data Group (PDG). The entry
was removed in later editions
when theoretical studies and partial-wave analyses
showed that the
bump around ${W=1685}$ MeV in the $\eta$-photoproduction cross
section on the neutron could also be due to
coupled-channel effects of known nucleon resonances 
\cite{Shklyar_07,Shyam_08}, effects from strangeness threshold
openings \cite{Doering_10} and cusps \cite{EtaMAID_19}, and
$S_{11}$ interferences
\cite{Anisovich_09, Anisovich_13,EPJA_S11_Interference}.
More insight will be gained from the ongoing measurements of polarization
observables, with first results for the observable $E$ already
available \cite{Witthauer_16,Witthauer_17,Witthauer_17b}.
This work revealed that a calculation of the Bonn-Gatchina (BnGa)
model \cite{EPJA_S11_Interference}
including a narrow $P_{11}$ resonance gave a slightly better
description of the experimental data than the model without
a narrow state. Due to the limited statistics,
the significance of this finding is still under
debate \cite{Anisovich_17}.

\subsection{Motivation for this work}
The latest claim of a $N_{\overline{10}}$ signature as a narrow
$N(1685)$ resonance is again made by Kuznetsov {\it et al.} using
GRAAL data in the $\gamma N\to \eta\pi N$
channels \cite{Kuznetsov_17}. Here, the $N_{\overline{10}}$
could be produced in the decay of a heavier $N$ or $\Delta$ resonance
$R$ via the emission of a pion $R\to \pi N_{\overline{10}}$ followed
by the decay $N_{\overline{10}}\to\eta N$. Therefore, in contrast to
the direct production potentially observed in $\gamma n\to \eta n$, the signal
could also be observed using a proton target despite the suppressed
photon coupling of the $N_{\overline{10}}$. Indeed, signals in
all four isospin channels of $\gamma N\to \eta\pi N$ are claimed in 
\cite{Kuznetsov_17}, whereas only the peak in
$\gamma p\to \eta\pi^0 p$ (measured on the free proton)
is statistically significant.	

The $\gamma N\to \eta\pi N$ reactions have been previously
measured at several experiments, such as
LNS \cite{Nakabayashi_06}, 
GRAAL \cite{Ajaka_08},
CBELSA/TAPS \cite{Horn_08,Gutz_11,Gutz_14} and
A2 at MAMI
\cite{Kashevarov_09,Kashevarov_10,Annand_15,Kaeser_16,Sokhoyan_18}.
The main findings were that for $E_\gamma < 1.6$ GeV, the
reaction is dominated by the excitation of the $\Delta(1700)3/2^-$
and the $\Delta(1940)3/2^-$ resonances with subsequent decays
to $\eta\Delta(1232)3/2^+$. Smaller contributions come from
the $\pi N(1535)1/2^-$ isobar and photoproduction of
the $a_0(980)$ meson. Evidence of a $N(1685)$ was not found
in any of these studies as statistics was either limited or the
event selection was not optimized to reveal a potential signal.
The goal of this work is therefore to measure $\gamma p\to \eta\pi^0 p$
and $\gamma p\to \eta\pi^+ n$
in order to look for any signs of a narrow structure around
${m(\eta N)\sim 1685}$ MeV. The available A2 data
for the energy range $E_{\gamma}$=1.43--1.58 GeV
have not been used so far to study these reactions.

\begin{figure}
\centering
\includegraphics[width=\columnwidth]{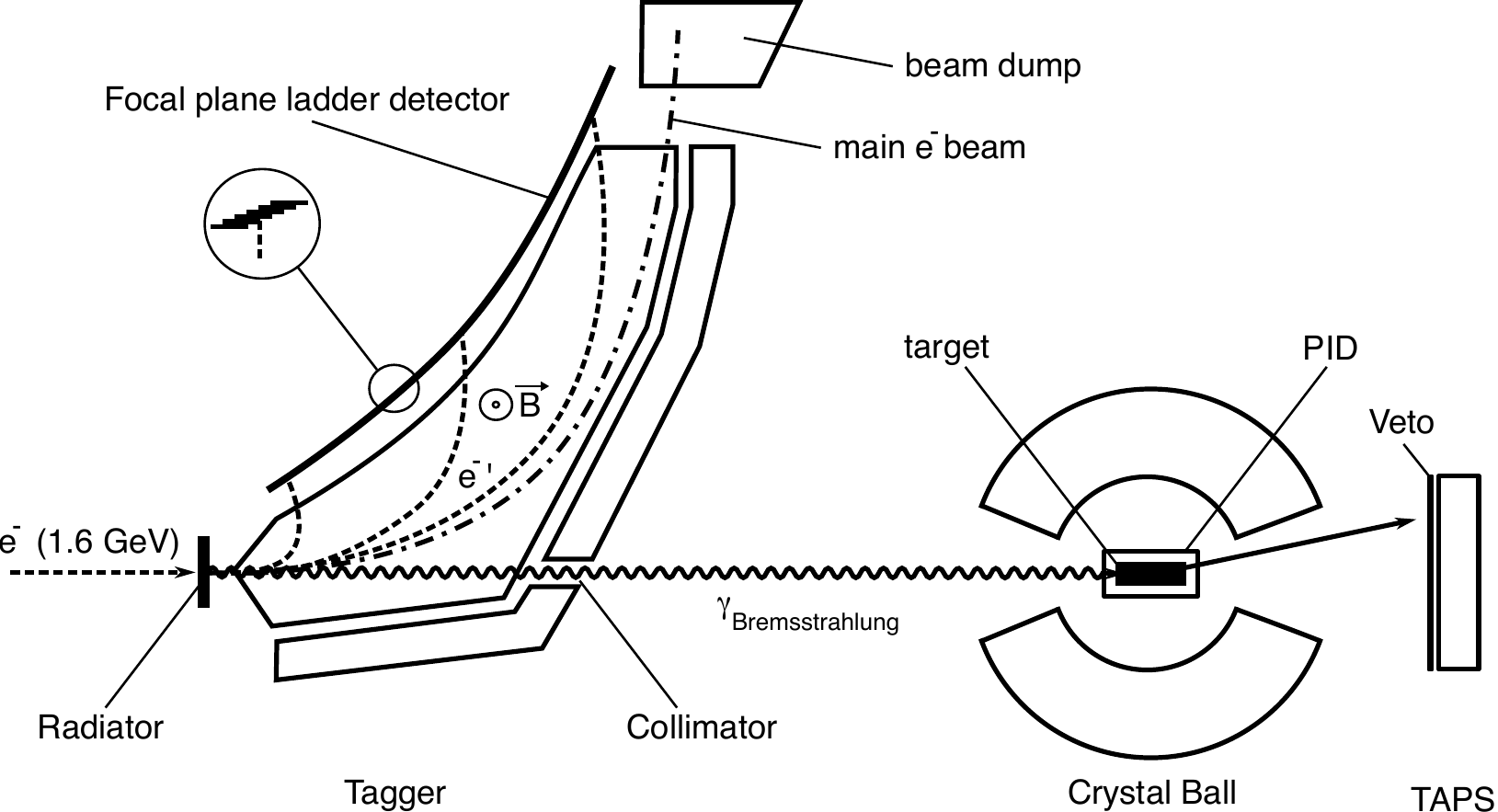}
\caption{Schematic view of the setup at the tagged-photon beam
experiment A2 at MAMI.}
\label{fig:setup}
\end{figure}

\section{Experimental Setup}
\label{sec:setup}
The experimental data were obtained at the tagged-photon 
beam facility A2 at the electron-accelerator
facility MAMI in Mainz, Germany. Out of three data sets of
similar size, one data set of 310 hours was used for this work.
Figure~\ref{fig:setup} shows an overview of the experimental setup.
In contrast to `standard' experiments, the Endpoint-Tagger (EPT)
device allowing to access photon-beam energies 
$E_{\gamma}$=1.43--1.58 GeV close
to the energy of the incoming 1.6 GeV electron beam was
installed.
The 10-cm liquid hydrogen target was installed in the center
of the Crystal Ball detector (CB) consisting of 672 NaI(Tl)
crystals, which
cover polar angles from 20--160 degrees with almost full
azimuthal acceptance. Particles with polar angles from 5--20 degrees 
were detected in the TAPS calorimeter wall installed 1.5 m downstream
from the target comprising 366 BaF$_2$ crystals. Thin plastic scintillator
detectors arranged as a barrel around the target in CB and in front of
individual TAPS crystals allowed charged-particle vetoing and particle
identification via a $dE/E$-analysis in both calorimeters. Particles
detected in TAPS could also be discriminated by time-of-flight and
pulse-shape analyses, the latter utilizing
the two scintillation-light components
in the BaF$_{2}$ crystals.
The experimental trigger required an energy deposition higher than
$\sim$550 MeV in CB as the primary goal of the EPT-experiments  
were studies involving the $\eta'$-meson. More details about the
setup can be found in \cite{Adlarson_15}.

\section{Data Analysis}
\label{sec:analysis}
The following reactions were analyzed in this work:
\\[1ex]
\begin{tabular}{l}
1) $\gamma p\to \eta\pi^0p$ with $\eta\to 2\gamma$\\
2) $\gamma p\to \eta\pi^+n$ with $\eta\to 2\gamma$\\
3) $\gamma p\to \eta\pi^+n$ with $\eta\to 3\pi^0\to 6\gamma$\\
\end{tabular}

\subsection{Event Selection}
\label{sec:event_sel}
As for now, exclusive analyses were performed, i.e., all particles
in the final state were required to be detected. The thresholds for
clusters in the CB and TAPS calorimeters were set to 20 MeV.
Using the detected particles, all combinations to form the
corresponding final states were checked with a series of cuts.
In addition, a kinematic fit was performed to optimize the measured
angles and energies of the particles using 4-momentum and
meson invariant-mass constraints.
Particle identification was performed with $dE/E$ (CB) and
time-of-flight (TAPS)
analyses for pion and neutron candidates in reactions 2) and 3).
BaF$_2$ pulse-shape cuts were used in all reactions to
separate photons from massive particles.
Next, cuts on the reaction kinematics were implemented. Neutral
pions and $\eta$-mesons were identified in the corresponding
$m(\gamma\gamma)$ invariant masses by applying $3\sigma$-cuts
around their nominal mass. All particles were required to
lie in the same reaction plane --- this was ensured by cutting on
the three azimuth-angle differences between the $p_ip_j$ and $p_k$
systems ($p\in\{\eta, \pi, N\}$).
A further cut was employed on the angular difference between the
detected and fitted direction of the nucleon.
The confidence level (CL) provided by the kinematic fit allowed
an additional cut (CL $> 2.7 \times 10^{-3}$) to reject
background events.
Having applied all cuts, the number of events with multiple
valid particle combinations were small. In order to account for
them in the signal extraction, large data sets of Monte-Carlo (MC)
simulations are needed. Therefore, for the current work only the
combination with the highest CL was kept for further analysis.

\begin{figure}
\centering
\includegraphics[width=\columnwidth]{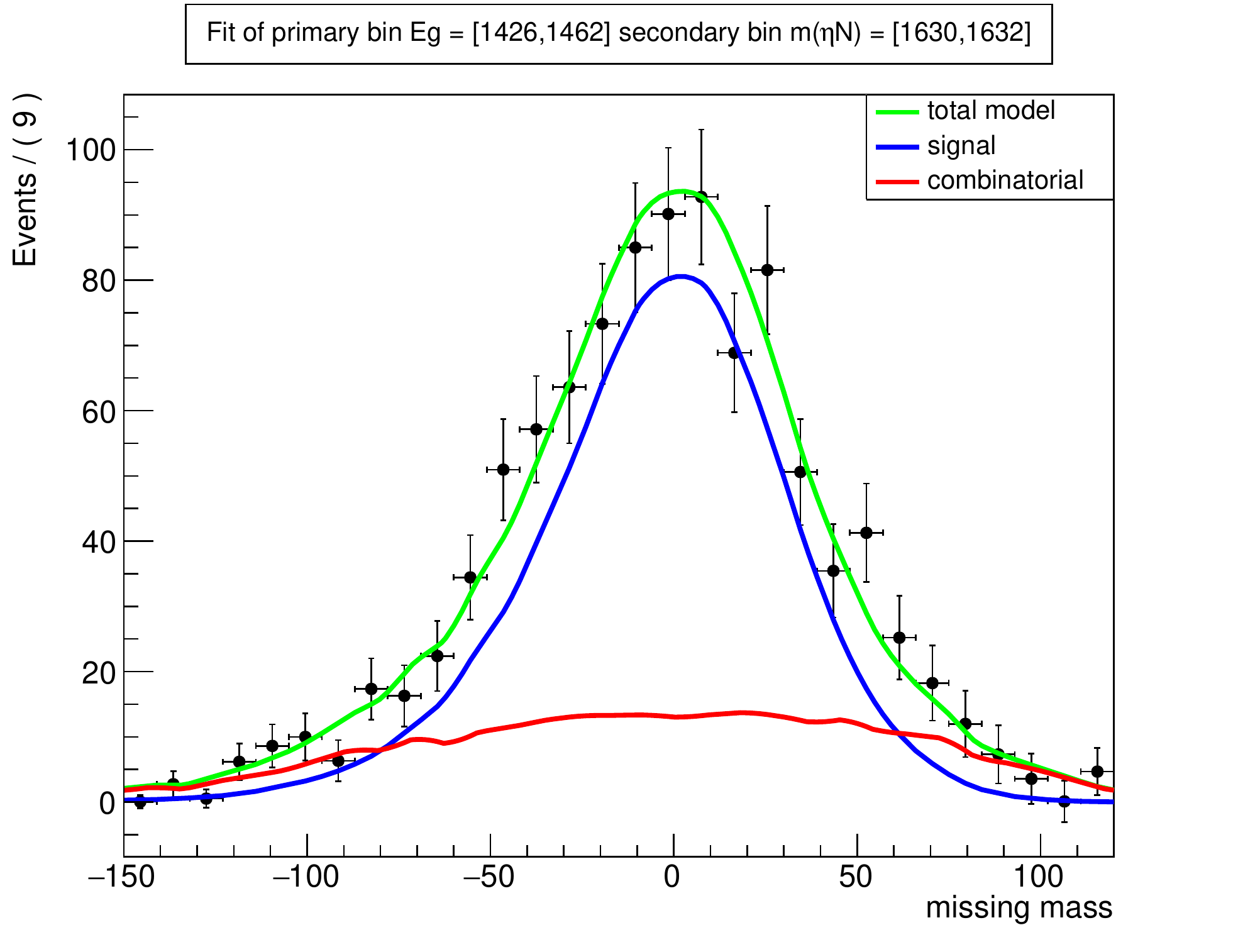}
\includegraphics[width=\columnwidth]{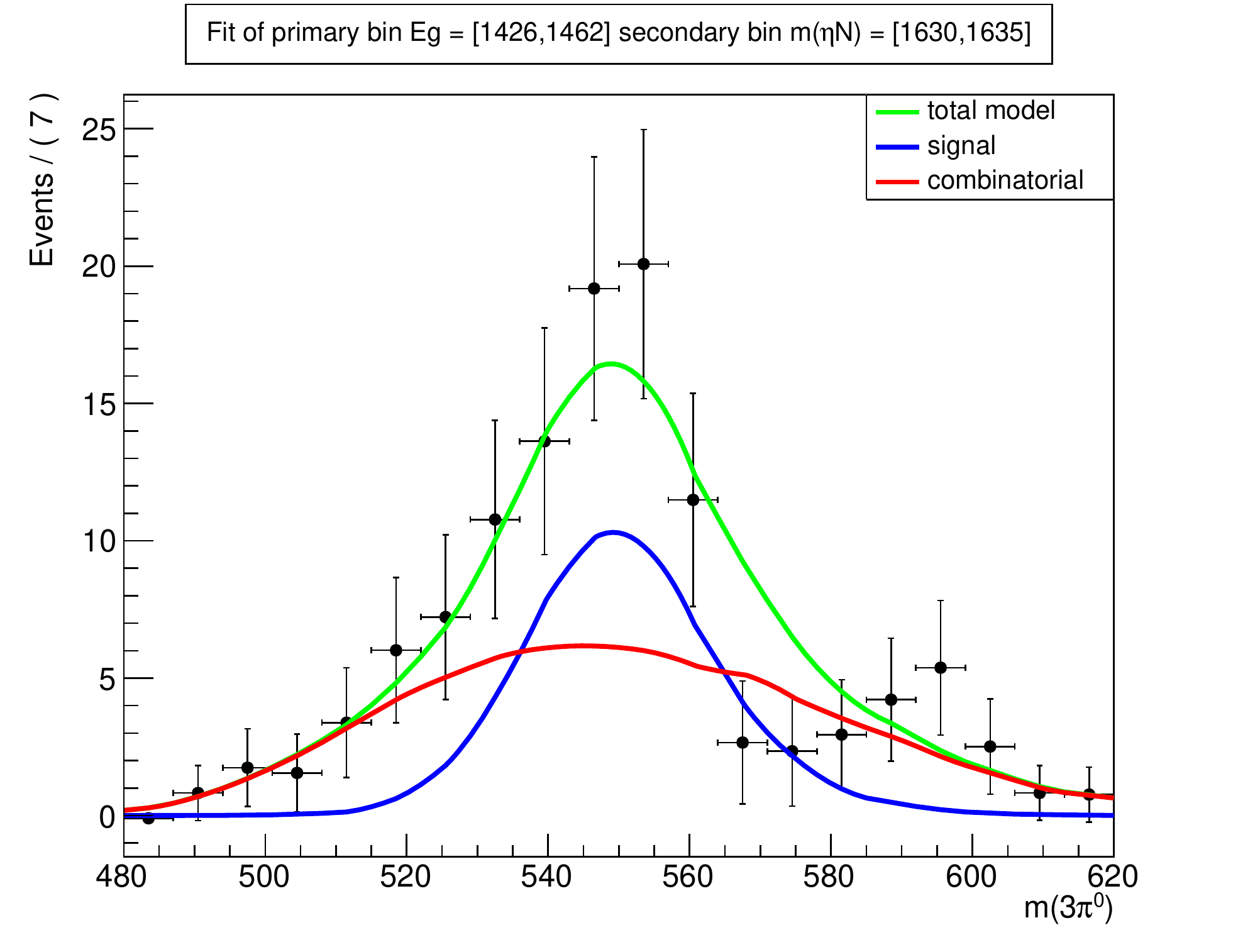}
\caption{Example unbinned likelihood fits for the same
$E_\gamma$, $m(\eta N)$-bin. Top: Fit of the $\eta\pi^0$-missing
mass in the analysis of ${\gamma p\to \eta\pi^0p}$
with ${\eta\to 2\gamma}$.
Bottom: Fit of the $m(3\pi^0)$-invariant mass in the analysis
of ${\gamma p\to \eta\pi^+n}$ with ${\eta\to 3\pi^0}$. Signal
and background contributions were obtained from MC simulations.}
\label{fig:splot_fits}
\end{figure}

\subsection{Signal Extraction}
The data were analyzed in bins of photon-beam energy $E_\gamma$
and $\eta$-nucleon invariant mass $m(\eta N)$. In order to remove any
background remaining after the previously discussed cuts, 
the $\eta\pi$-missing mass (in case of reaction 1) and 2),
nucleon mass subtracted) or
the $m(3\pi^0)$-invariant mass (in case of reaction 3)) were
fitted with signal and background contributions obtained from MC
simulations using a Geant4-based \cite{Geant4}
model of the A2 experiment.
The reason for the choice of different fit variables is
the fact that very little background is present in analysis 3)
and the $m(3\pi^0)$-invariant mass provided more stable and
unambiguous fits compared to the $\eta\pi$-missing mass.
Example fits of both types are shown in figure~\ref{fig:splot_fits}.
Unbinned likelihood fits were performed which allowed to make
use of the sPlot-technique \cite{Pivk:2005ca} to obtain weights
for the signal and background contribution on an event-by-event
basis. This enables to unfold signal and background in any variable
that is not correlated with the fit variable.

\subsection{Van Hove Plots and Longitudinal Phase-Space}
\label{sec:van_hove}
The use of longitudinal phase-space
allows to reduce the phase-space dimensionality
of multi-particle final states
by assuming that the transverse momenta can be neglected
in high-energy particle collisions \cite{VanHove_69}.
In case of a three-particle final state, the longitudinal phase-space
has two dimensions and Van Hove plots containing six sectors can
be constructed. The sectors represent the possible 
combinations of the longitudinal momenta of all particles, i.e.,
whether they are going forward (direction of beam)
or backward (direction of target) in the center-of-mass
frame. As different reaction mechanisms will populate different sectors,
Van Hove plots can be used to disentangle different processes.

\begin{figure}
\centering
\includegraphics[width=0.9\columnwidth]{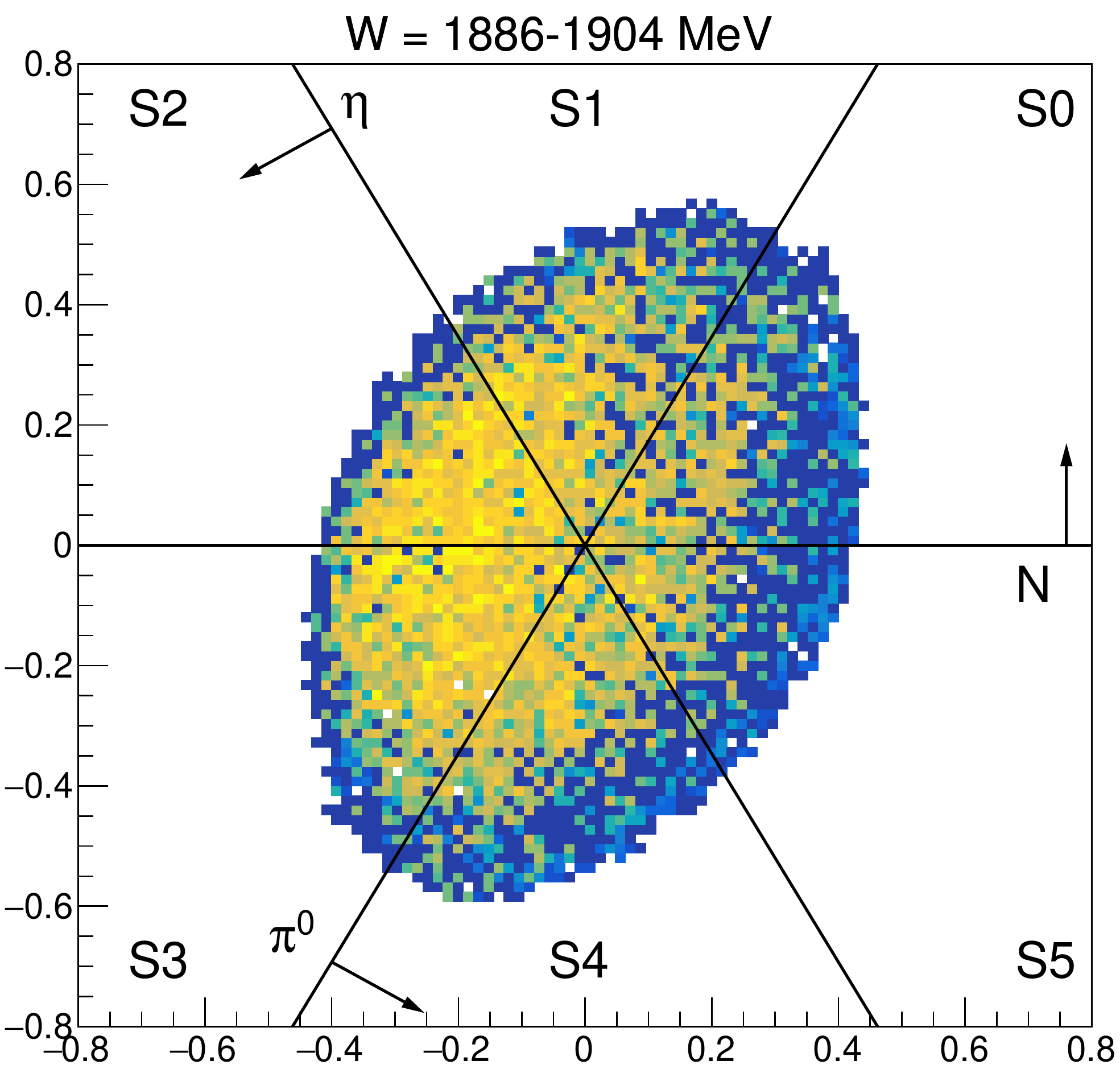}
\caption{Van Hove plot for $\gamma p\to\eta\pi^0p$
in the energy range $E_\gamma$=1426--1462 MeV. The six sectors
S0 to S5 represent different combinations of forward or backward
going particles (see text).}
\label{fig:lps}
\end{figure}

The aim of the work performed here was to investigate the possibility
of using Van Hove plots to separate the potential production
of the $\pi N(1685)$ isobar from the dominating $\eta\Delta(1232)$
isobar.
The main caveat is that the in the studied energy range
the transverse momenta cannot be neglected as required by the
longitudinal phase-space approximation. Nevertheless, a partial
separation could already lead to a sufficient improvement of the
signal-to-background ratio.

Figure~\ref{fig:lps} shows the Van Hove plot for the first energy
bin in the $\gamma p\to\eta\pi^0p$ analysis. The sector with the
largest number of events is S3, consistent with the
dominating production of
the $\eta\Delta(1232)$ isobar, with a forward going $\eta$-meson
and $\pi^0p$ going backward. Similarly, a possible $\pi N(1685)$
isobar is expected to be predominantly located in sector S5
assuming a production mechanism which yields a backward-going $N(1685)$ and a 
forward-going $\pi$-meson.
As mentioned before, due to the relatively low beam energy,
the locii for the different production mechanisms are not
confined to one sector and strongly overlap.

\section{Preliminary Results}
\label{sec:results}
\begin{figure*}
\centering
\includegraphics[width=0.99\textwidth]{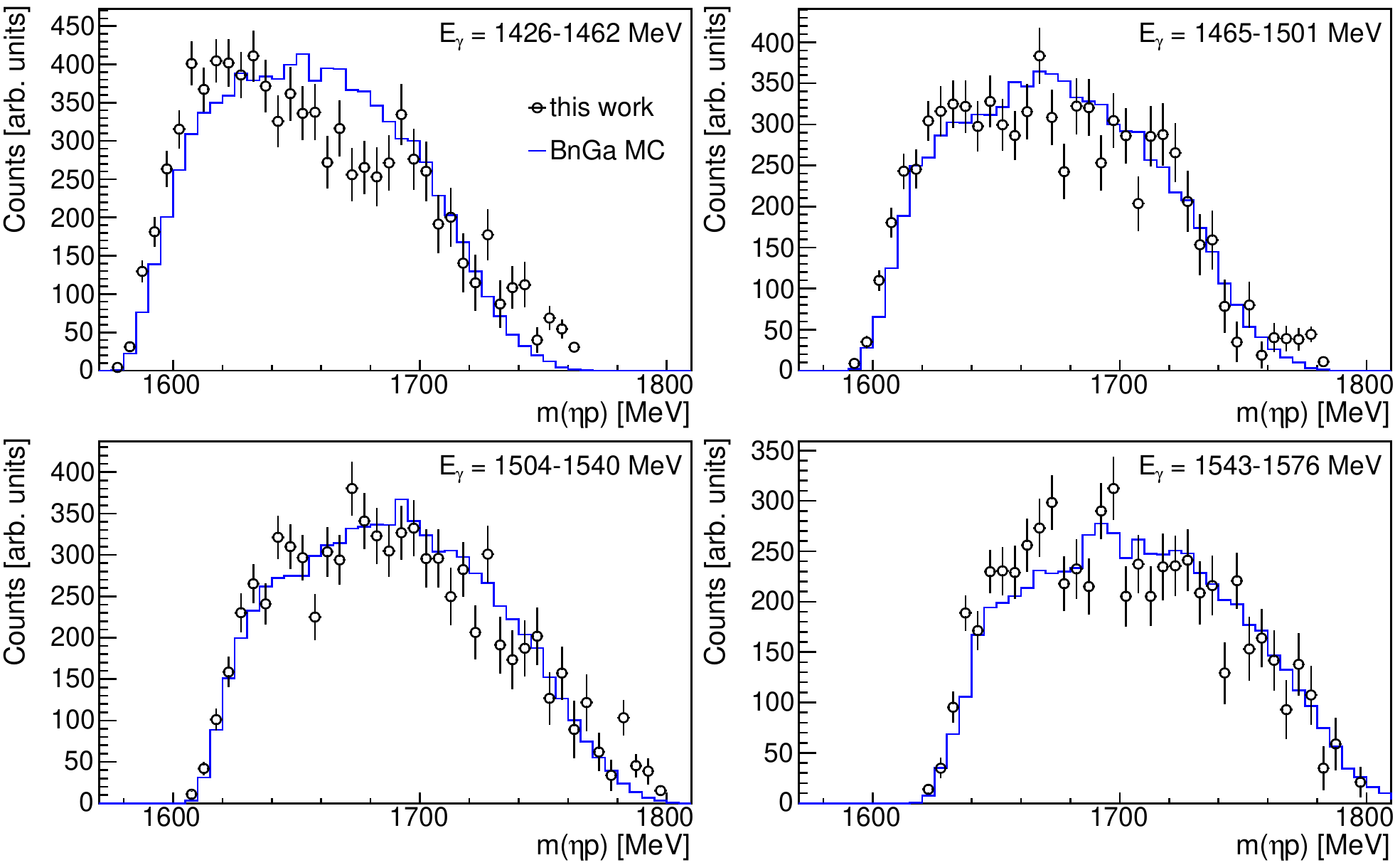}
\caption{Invariant mass $m(\eta p)$ of $\gamma p\to \eta\pi^0p$ for
four bins of photon-beam energy with $m(\pi^0p) <$ 1180 MeV
(no acceptance correction).
The data (black circles) are compared to a phase-space MC
simulation (blue curves, normalized by integral ratio) using event
weights from the BnGa partial-wave analysis of \cite{Gutz_14}.}
\label{fig:m_eta_p_1180}
\end{figure*}

\begin{figure*}
\centering
\includegraphics[width=0.99\textwidth]{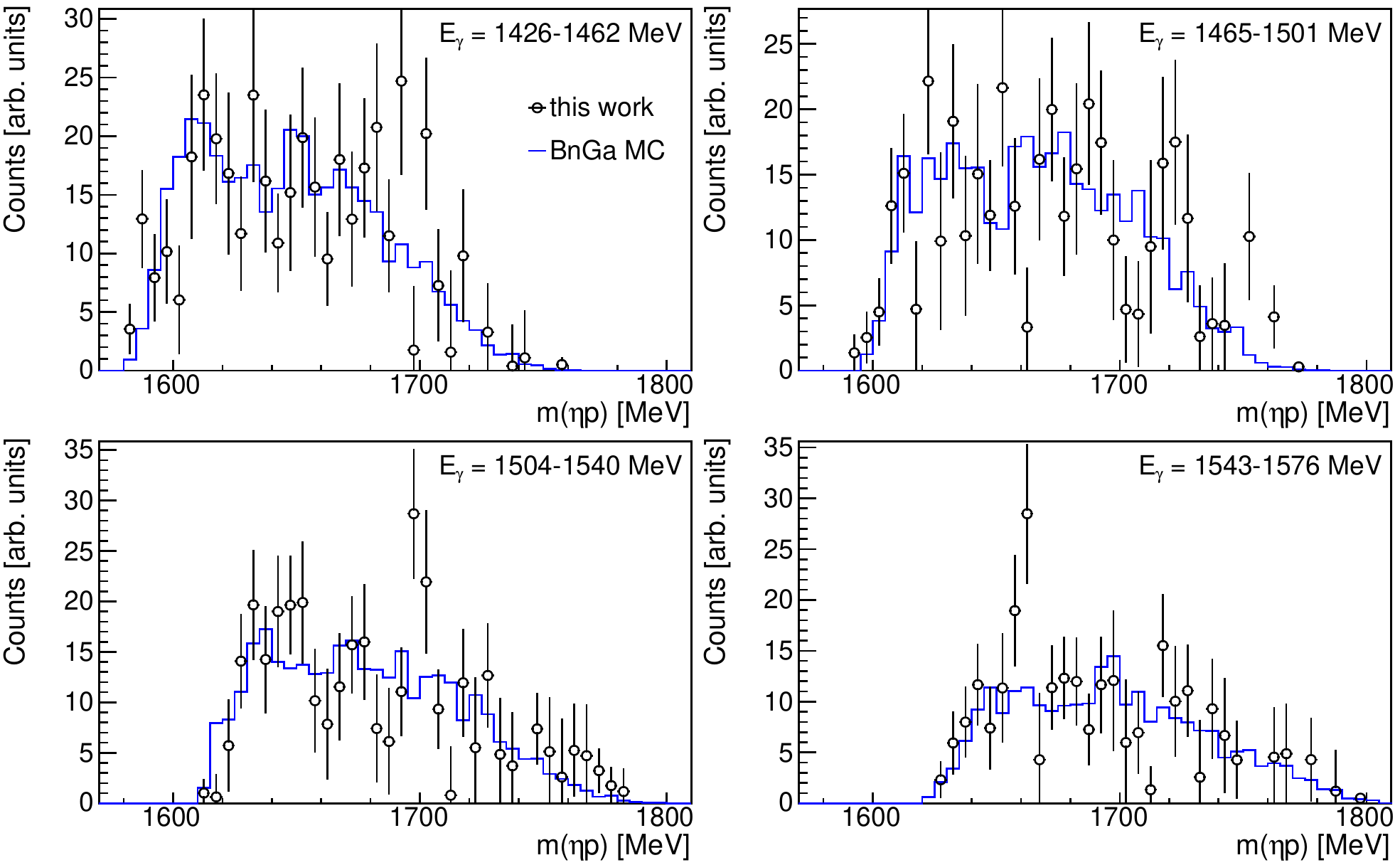}
\caption{Invariant mass $m(\eta p)$ of $\gamma p\to \eta\pi^0p$ for
four bins of photon-beam energy with $m(\pi^0p) <$ 1180 MeV for
events in the Van Hove plot-sector S5
(no acceptance correction).
The data (black circles) are compared to a phase-space MC
simulation (blue curves, normalized by integral ratio) using event
weights from the BnGa partial-wave analysis of \cite{Gutz_14}.}
\label{fig:m_eta_p_1180_sec5}
\end{figure*}

Preliminary results of the observable of interest, the
$m(\eta N)$ invariant mass, are shown in
figures~\ref{fig:m_eta_p}--\ref{fig:m_eta_p_1180_sec5} and will
be discussed in the following.

\subsection{Uncut Distributions}
\begin{figure*}
\centering
\includegraphics[width=\textwidth]{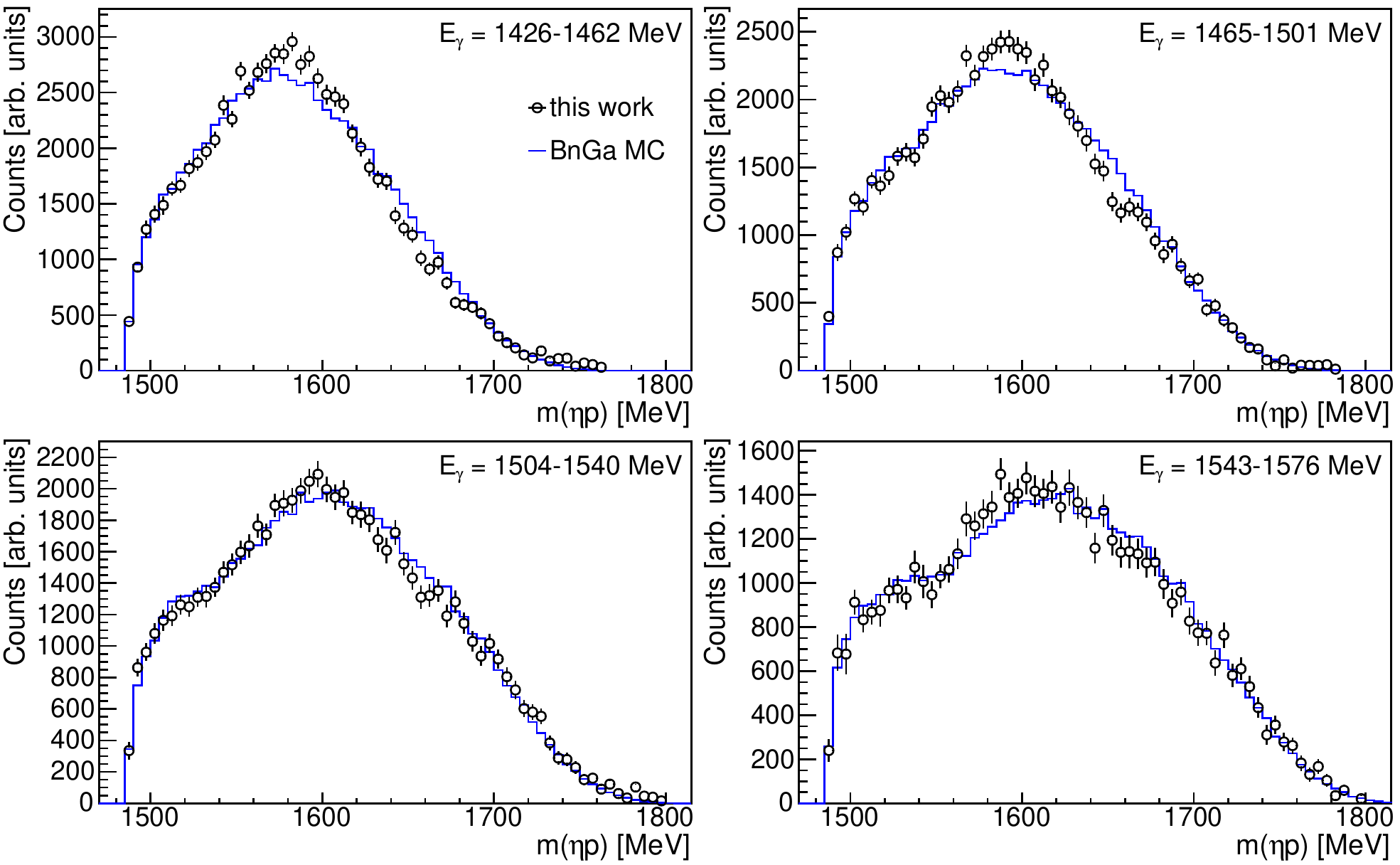}
\caption{Invariant mass $m(\eta p)$ of $\gamma p\to \eta\pi^0p$ for
four bins of photon-beam energy (no acceptance correction).
The data (black circles) are compared to a phase-space MC
simulation (blue curves, normalized by integral ratio) using event
weights from the BnGa partial-wave analysis of \cite{Gutz_14}.}
\label{fig:m_eta_p}
\end{figure*}

First, the invariant mass distributions were studied
using the standard analysis cuts discussed in
section~\ref{sec:event_sel}. In case of the
$\gamma p\to\eta\pi^0p$ reaction, the results 
shown in figure~\ref{fig:m_eta_p} are compared
to the distributions obtained from phase-space MC simulations using
event weights from the BnGa partial-wave analysis of \cite{Gutz_14}, 
which were scaled to the data using the ratio of the distribution
integrals. No acceptance correction has been applied here.
There is an overall good agreement between the two distributions
with a small discrepancy at the maximum, where the data shows
a slightly sharper shape. No excess around 1685 MeV is observed.
Only the $\eta\to 2\gamma$ decay has been analyzed so far, as the
$\eta\to 3\pi^0$ decay leads to a total number of nine particles
in the final state, which will degrade the
detection efficiency resulting in a smaller data set.

\begin{figure*}
\centering
\includegraphics[width=\textwidth]{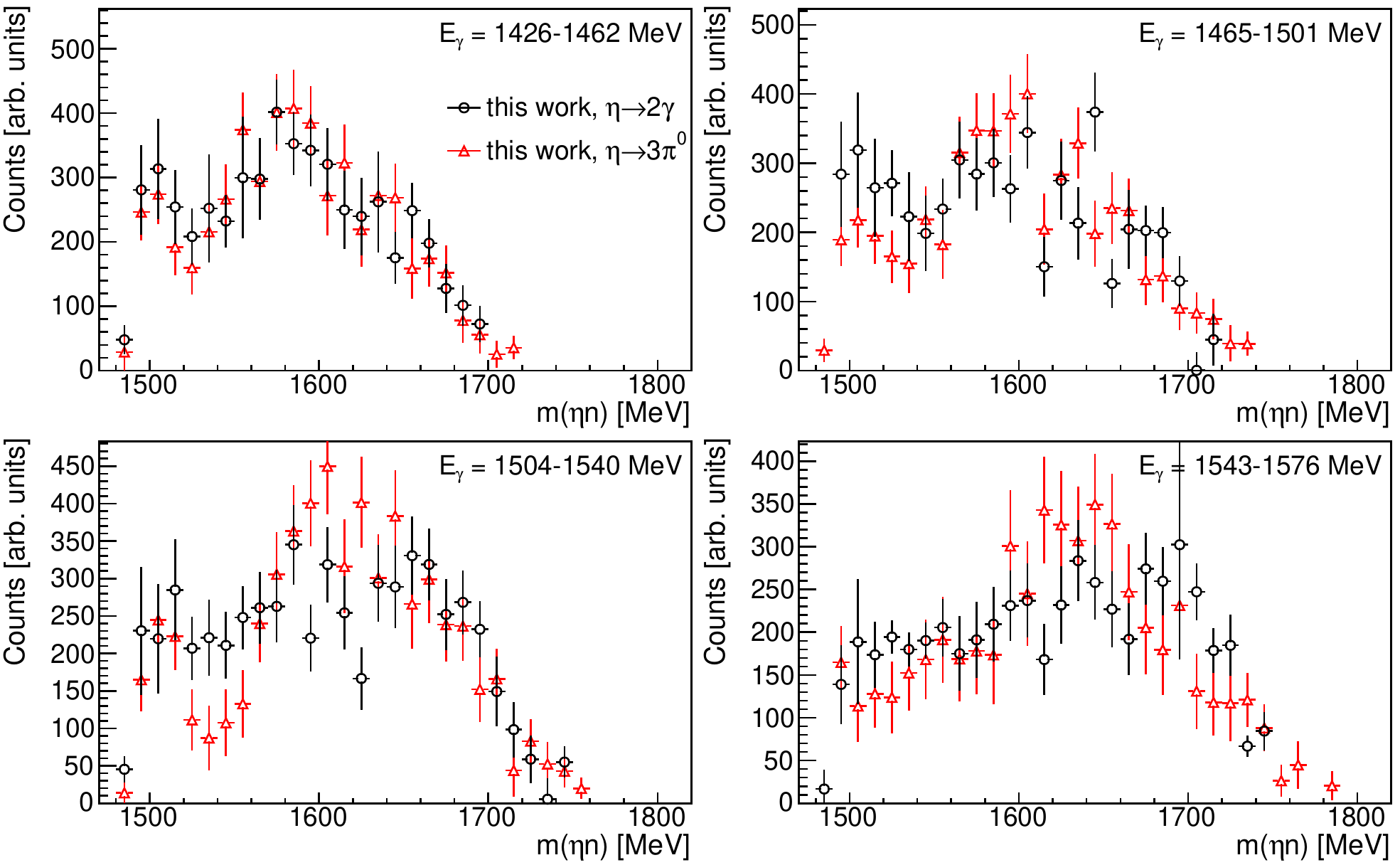}
\caption{Invariant mass $m(\eta n)$ of $\gamma p\to \eta\pi^+n$ for
four bins of photon-beam energy (no acceptance correction).
Two data samples using the $\eta\to 2\gamma$ decay (black circles)
and the $\eta\to 3\pi^0$ decay (red triangles, normalized by
integral ratio) have been analyzed.}
\label{fig:m_eta_n}
\end{figure*}

Figure~\ref{fig:m_eta_n} shows the $\eta n$-invariant mass
for $\gamma p\to\eta\pi^+n$. Both decays of the $\eta$-meson
were analyzed and are compared to each other by scaling
the $\eta\to 3\pi^0$ results to the results of the $\eta\to 2\gamma$
data sample. Despite the lack of an individual efficiency correction,
the overall features of the distributions are rather similar.
Statistically significant finer structures are hard to identify
due to the limited statistics. The main reasons for the worse
statistical quality are the lower ($\sim$$1/3$) detection
efficiency for neutrons compared to protons and the fact that
the experimental trigger required a high total energy 
($\sim$550 MeV) deposited in the CB detector. Since neutrons deposit
on average less energies than protons, and the $\pi^+$ does not
contribute to the detected energy with its whole rest mass
in contrast to the $\pi^0$, the trigger is less efficient resulting
in a smaller data set.

\subsection{Enhancing a Potential $\boldsymbol{N(1685)}$-Signal}
In order to enhance a possible signal of a narrow $N(1685)$ state,
additional analysis cuts have to be applied. At the moment, only the
${\gamma p\to \eta\pi^0p}$ reaction was studied regarding further
cuts as statistics of the current ${\gamma p\to \eta\pi^+n}$
data sample is too poor. This is expected to improve when all available
data will have been analyzed.

In order to suppress contributions from the $\eta\Delta(1232)$ isobar,
a cut requiring $m(\pi^0p) <$ 1180 MeV was applied. The resulting
$m(\eta p)$ distributions are shown in figure~\ref{fig:m_eta_p_1180}.
The data are again compared to a MC simulation weighted with the
BnGa-model that was scaled to the data by the ratio of integrals.
In bins two to four, the overall agreement is quite good considering
the statistical uncertainties of both data and model. The statistical
errors of the model are not shown directly but fluctuations are clearly
visible. These fluctuations will be removed by generating a larger
MC data sample in the final analysis.
In the first bin, there are some noticeable
deviations of the data from the model: 
For $m(\eta p)$=1640--1680 MeV, the model seems to overestimate
the data, which form a peak-like structure around 1690 MeV.
The statistical significance of these features is low, though.
Also, the scaling of the model to the data will need to be implemented
using absolute normalizations in the final analysis rather than
using a factor that averages over the whole distributions.

Figure~\ref{fig:m_eta_p_1180_sec5} shows the $m(\eta p)$ distributions
including the $m(\pi^0p) <$ 1180 MeV cut and requiring in addition
that the event belongs to sector S5 in the Van Hove plot. As discussed
in section \ref{sec:van_hove}, this is expected to be the
predominant sector for the potential $\pi N(1685)$ isobar.
Statistics is obviously further reduced and fluctuations are also
significant in the BnGa-model. Nevertheless, overall the data and
model distribution agree rather well. Finer structures cannot be
identified due to the limited statistics, although there a few
bins with more entries around 1700 MeV, again not statistically
significant.

\section{Summary and Outlook}
\label{sec:summary}
The reactions $\gamma p\to\eta\pi^0p$ and
$\gamma p\to\eta\pi^+n$ in the energy range
$E_\gamma$=1.43--1.58 GeV have been analyzed using data
obtained at the A2 at MAMI experiment. The goal of this work
was to check recent claims \cite{Kuznetsov_17} of a signature
of the $N(1685)$-antidecuplet baryon in the $m(\eta p)$ and
$m(\eta n)$ invariant masses of the reactions
$\gamma N\to\eta\pi N$.

For $\gamma p\to\eta\pi^0p$, our data show a good agreement
with previous measurements represented by the BnGa partial-wave
analysis \cite{Gutz_14}. When a cut requesting $m(\pi^0p) <$1180 MeV
to suppress the dominating $\eta\Delta(1232)$ background is applied,
there are some deviations between the data and the model around
the region of interest for photon-beam energies
$E_\gamma$=1426--1462 MeV.
The statistical significance of these
deviations is small, nevertheless this finding needs to 
be investigated further by increasing the statistical quality
of both data and model. Having more data will also allow
to get better statistics when using an additional cut on
the preferred signal sector in the Van Hove plot, which could
help isolating a small $\pi N(1685)$ contribution to the reaction.

The $\gamma p\to\eta\pi^+n$ analysis currently lacks the statistical
quality for a refined analysis regarding a narrow $N(1685)$.
All of the available data need to be analyzed in combination
with an optimized event selection to come to a final conclusion here.

Next steps will include optimizations in the analysis
($\eta\to 3\pi^0$ decay for $\eta\pi^0p$ final state, cuts, test of
inclusive analyses), the use of all available data,
the generation of a larger MC data sample using
the BnGa-model, and acceptance correction with absolute normalization
and comparison with the Mainz model \cite{Sokhoyan_18}.

\bibliography{references}

\end{document}